\documentclass{ws-ijmpa}

\begin{document}

%
\catchline{}{}{}{}{}
%

\title{Towards the understanding of the meson spectra}

\author{J. Vijande, F. Fern\'andez, A. Valcarce}

\address{Nuclear Physics Group,\\
University of Salamanca,\\
Plaza de la Merced s/n, E-37008 Salamanca, Spain}

\maketitle

\pub{Received (Day Month Year)}{Revised (Day Month Year)}

\begin{abstract}
We present a quark-quark interaction
for the complete study of the meson spectra, from the light
to the heavy sector. We compare the quark model predictions against
well-established $q\bar q$ experimental data.
This allows to identify discrepancies between quark model results and
experiment that may signal physics beyond conventional hadron spectroscopy.
\end{abstract}

\section{SU(3) chiral constituent quark model}

Since the origin of the quark model hadrons have been considered as
bound states of constituent (massive) quarks. The constituent quark mass
appears as a consequence of the spontaneous
breaking of the original $SU(3)_{L}\otimes SU(3)_{R}$ chiral symmetry at
some momentum scale. This symmetry breaking also generates Goldstone modes
that are exchanged between the constituent quarks.
Explicit expressions and details of these potentials can 
be found elsewhere\cite{vij1}. 
In the heavy quark sector, chiral symmetry is
explicitly broken and therefore Goldstone modes do not appear.

Above the chiral symmetry breaking scale quarks still interact through
gluon exchanges. Following de R\'{u}jula {\it et al.}\cite{ruju}
the one-gluon-exchange (OGE) interaction can be taken as a 
standard color Fermi-Breit potential.
To obtain a unified description of light, strange and 
heavy mesons a running strong coupling constant has to be used\cite{godf}. 
The perturbative expression for $\alpha_s(Q^2)$ diverges 
when $Q\rightarrow\Lambda_{QCD}$ and therefore the coupling
constant has to be frozen at low energies. 
This behavior is parametrized by
means of an effective scale dependent strong 
coupling constant\cite{vij1,coup}.
The spin-spin interaction of the OGE presents a contact term that has to be 
regularized in order to avoid an unbound spectrum from below. 
For a coulombic system
the size scales with the reduced mass, what suggests to use a
flavor-dependent regularization $r_0(\mu)={{\hat r_0}/{\mu}}$\cite{vij1}.
Moreover the Schr\"{o}dinger equation cannot be solved numerically 
for potentials containing $1/r^3$ terms.
This is why the noncentral terms of the
OGE are also regularized.

Finally, potential models have to include other nonperturbative property of QCD, confinement.
Lattice QCD studies show that $q\overline{q}$ systems 
are well reproduced at short distances by a linear potential that
it is screened at large distances due to pair creation\cite{bali}.
One important question is the covariance property of confinement. While the
spin-orbit splittings in heavy quark systems 
suggest a scalar confining potential\cite{luca}, 
Szczepaniak and Swanson \cite{szcz} 
showed that the Dirac structure of confinement is of vector nature in
the heavy quark limit of QCD. Therefore, we write 
the confining interaction 
as an arbitrary combination of scalar and vector terms.

Using this model we have studied more than 110 states
from the light to the heavy sector\cite{vij1}, obtaining a rather good description.
We have analyzed in detail several states reported in the charm sector which do not
seem to fit into a $q\bar q$ structure. The same applies for the light
scalar sector which does not seem possible to be 
described using only $q\bar q$ states.

\section{New results on charmonium physics}

During the last year several new results of the $\eta_c(2S)$ mass have been
reported. They are significantly larger than most predictions of constituent
quark models and the previous experimental value of the PDG: 
$M[\eta_c(2S)]=3594 \pm 5$ MeV\cite{bel1}.
We find for this state an energy of 3627 MeV,
close to the new experimental value quoted by the PDG, $3637.7\pm4.4$ MeV.

BaBar has reported a narrow state near 2317 MeV called $D_{sJ}^{\ast }(2317)$\cite{baba}. This
state has been confirmed by CLEO\cite{cleo} together with another
possible resonance around 2460 MeV. Both experiments interpret 
these resonances as $J^P=0^+$ and $1^+$ states. 
The most striking aspect of these two resonances is that their masses are
much lower than expected. Our results are 2470 MeV for the $D_{sJ}^{\ast
}(2317)$ and 2550 for the $D_{sJ}^{\ast }(2460)$. The other open-charm
states agree reasonably well with the values of the PDG, but 
the two new $D_{s}$ states reported by Belle and CLEO do not fit into a $q\bar
q$ scheme.

The most recent state discovered in the charm sector is the $X(3872)$, 
which was reported by Belle\cite{chi} with a mass of $3872.0\pm0.6\pm0.5$ MeV and
confirmed by CDF\cite{chi2}. One of its most interesting features is that its energy is within the
error bars of the $D^0D^{0*}$ threshold, $3871.5\pm0.5$ MeV. Due to its
experimental decays the most probable assignment for this state would be an $L=2$ $c\bar c$ state, 
however most of the quark models predict a somewhat lower mass\cite{chi3}. Our model gives 3790 MeV 
for the $2^{--}$, 3803 MeV for the $3^{--}$ and 3793 MeV for the $2^{-+}$, which are too low to
be identified with it. Another possibility is that this state could be an
excited $1^{++}$ $P-$wave\cite{chi4}, but again our prediction, 3913 MeV, does not match
the experimental energy.

\section{The light scalar sector}

With respect to the isovector state, we find a candidate
for the $a_{0}(980)$, the $^3P_0$ member of the lowest $^3P_J$ 
isovector multiplet, with and energy of 983 MeV. 
However, in spite of the correct description of the mass 
of the $a_{0}(980)$, the model predicts a pure light-quark content,
what seems to contradict some of the observed decays. 
In the case of the isoscalar states, one finds a candidate for the $f_{0}(600)$
with a mass of 413 MeV, in the lower limit of the
experimental error bar, but the $f_{0}(980)$ cannot be found for any
combination of the parameters of the model. 
Concerning the $I=1/2$ sector our model predicts a mass
for the lowest $0^{++}$ state 200 MeV greater than the 
$a_{0}(980)$ mass. Therefore the $\kappa (800)$
cannot be explained as a $q\bar q$ pair.

It is worth to notice that similar conclusions has been achieved using
approaches as different as chiral perturbation theory\cite{pela} or
an extended Nambu-Jona-Lasinio model in an improved 
ladder approximation of the Bethe-Salpeter equation\cite{umek}. This seems to indicate that
other corrections would not improve the situation and the conclusions remain model
independent.               

\section*{Acknowledgments}
This work has been partially funded by Ministerio 
de Ciencia y Tecnolog{\'{\i}}a under Contract No. BFM2001-3563, 
by Junta de Castilla y Le\'{o}n under Contract No. SA-104/04.

\end{document}